# Spin Echo Studies on Cellular Water

D. C. CHANG[1,2], C. F. HAZLEWOOD[1], B. L. NICHOLS[1], H. E. RORSCHACH[2]

[1]Departments of Pediatrics and Physiology, Baylor College of Medicine, Texas Children's Hospital, and [2]Department of Physics, Rice University, Houston, Texas 77025

Previous studies indicated that the physical state of cellular water could be significantly different from pure liquid water. To experimentally investigate this possibility, we conducted a series of spin-echo NMR measurements on water protons in rat skeletal muscle. Our result indicated that the spin-lattice relaxation time and the spin-spin relaxation time of cellular water protons are both significantly shorter than that of pure water (by 4.3-fold and 34-fold, respectively). Furthermore, the spin diffusion coefficient of water proton is almost 1/2 of that of pure water. These data suggest that cellular water is in a more ordered state in comparison to pure water.

High resolution nuclear magnetic resonance (NMR) studies on rat skeletal muscle have recently shown that cellular water produces an absorption line almost ten times as broad as the line width of pure water[1]. This suggests that these water molecules are at least partly bounded such that the proton correlation time is increased. Similar observations have also been made for deuterated water in skeletal muscle[2] and for $H_2O$ in other tissues[3–5]. Our results further substantiate the notion that ordered water exists in biological tissue and rules out several criticisms of this interpretation.

To obtain more precise and detailed information about the cellular water, we have complemented the high resolution studies. The spin-lattice relaxation time ($T_1$), the spin-spin relaxation time ($T_2$) and diffusion coefficient ($D$) were measured by spin-echo NMR with procedures given elsewhere[6,7]. All experiments were performed at room temperature and the results are summarized in Table 1. Pure water in identical experimental conditions gave values for $T_1$, $T_2$, and $D$ very close to earlier published data[8,9]. We find $T_1/T_2 = 1.85$ for pure water in agreement with the observations of Meiboom et al.[8].

Table 1  Measurements of Relaxation Times and Diffusion Coefficients

|  | $T_1$ (s) | $T_2$ (s) | $D \times 10^{-5}$ (cm$^2$/s) |
|---|---|---|---|
| Pure water | 3.09±0.15 (4) | 1.52±0.093 (9) | 2.78±0.035 (11) |
| Skeletal muscle | 0.72±0.05 (6) | 0.045±0.002 (12) | 1.43±0.07 (5) |

All values are mean ±standard error of the mean. The numbers in parentheses represent the number of samples analysed.

These measurements on mammalian muscle water agree with the findings of the previous high resolution[1] and pulsed[3, 4] NMR studies. The diffusion coefficient of muscle water was also in qualitative agreement with that determined for a variety of biological tissues[4]. The observations suggest that the change of water properties (in the NMR sense) in the cellular environment is a universal phenomenon which is independent of species and tissue. The interpretation of the NMR results, however, is still controversial.

Some investigators favour the interpretation that the dramatic shortening of the proton relaxation times implies a structural change of the cellular water; water molecules inside the tissue are more ordered than those of ordinary water[1,2], but others do not agree with this interpretation. Hansen and Lawson[10] argued that the high resolution NMR line broadening might be caused by the diffusion of free water through microscopic magnetic field inhomogeneities. The pulsed NMR results are not consistent with this broadening mechanism. First, the shortened values of $T_1$ for cellular water cannot be explained by diffusion in an inhomogeneous field. Second, if this mechanism is to be effective in reducing the apparent spin-spin relaxation time, then an unreasonably large value for the local field inhomogeneity must be assumed. It was shown by Carr and Purcell[6,7] that the measured $T_2$ in an infinite sample is given by

$$\frac{1}{T_{2(effective)}} = \frac{1}{T_2} + \frac{1}{3} \gamma^2 G^2 \tau^2 D \qquad (1)$$

Here, $D$ is the diffusion coefficient, $\gamma$ is the gyromagnetic ratio, $G$ is the magnetic field gradient and $\tau$ is the time between the 90° pulse and the 180° pulse. In a bounded system the effective $D$ is smaller than the true diffusion coefficient[11-13]. Suppose we assume an upper limit for $D$ and set it equal to the value for pure water. Then for a typical $\tau \sim 0.03$ s and $T_2$ measured $\sim 0.04$ s, a value of $G$=2.2 gauss/cm would be required, which is more than twenty times the natural sample magnetic field inhomogeneity as determined from the width of the echo signal. Furthermore, we have measured the value of $T_2$ for muscle water for various values of $\tau$ from 13 ms to 36 ms. The values of $1/T_{2(effective)}$ were unchanged within experimental error (Fig. 1). If the relaxation is the result of molecular diffusion through a magnetic field gradient, the reciprocal of the measured $T_2$ would be a linear function of $\tau^2$ (ref. 10 and equation (1)). This dependence is not observed in our experiments.

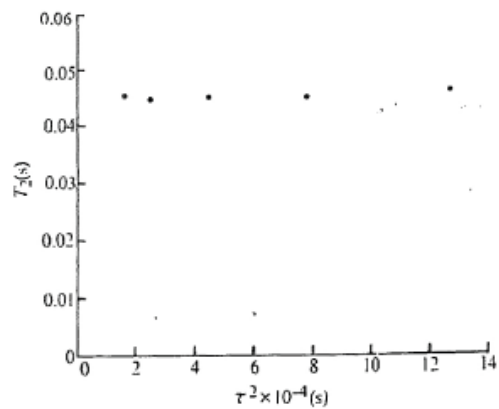

Fig. 1 Spin-spin relaxation time ($T_2$) of proton in muscle water as a function of observation time ($\tau$).

Other objections to the ordered water interpretation have also been advanced. For example, Glasel has pointed out[14] that the exchange of solvent species between the bulk phase, and that which exists near suspended matter is sufficient to cause line broadening, and no ordering or disordering of the solvent must be postulated. To support this argument, Glasel cites experiments on water in systems packed with glass beads to show that the relaxation times of protons and deuterons change as the surface-to-volume ratio varies. We feel, however, that the results of these experiments do not necessarily negate the ordered water interpretation.

Glass is hydrophilic and has a strong Van der Waals interaction with water, which can be easily seen from the capillary effect. Hori's freezing experiments[15] showed that a thin layer (greater than 1 μm) of water held between two glass surfaces is highly structured. Therefore, it is clear that water molecules must be ordered near the surface of the glass beads used by Glasel[14]. As the radius of the glass bead is reduced, the surface area of the glass-water interface, as well as the percentage of structured water, increases. This increase in water order can cause the relaxation times to be reduced. Glasel's experiments therefore do not disprove the ordered water interpretation, but support it. In addition, evidence for the existence of a highly structured water near interfaces is reported elsewhere for physical[15-20] and biological systems[21-23].

Parts of this work were supported by the Medical Research Foundation of Texas, the Robert A. Welch Foundation, and grants from the US Public Health Service. We thank Mr. David Harris and Miss Marie-Pierre Barthe for technical assistance and Mrs. Betty E. Perronne for assistance in the preparation of this manuscript.

Received April 5; revised July 6, 1971.

1. Hazlewood, C. F., Nichols, B. L., and Chamberlain, N. F., *Nature*, **222**, 747 (1969).
2. Cope, F. W., *Biophys. J.*, **9**, 303 (1969).
3. Bratton, C. B., Hopkins, A. L., and Weinberg, J. W., *Science*, **147**, 738 (1968).
4. Abetsedarskaya, L. A., Miftakhutdinova, F. G., and Fedorov, V. D., *Biofizika*, **13**, 630 (1968); translation in *Biophysics*, **13**, 750 (1968).
5. Swift, T. J., and Fritz, O. G., *Biophys. J.*, **9**, 54 (1969).
6. Carr, H. Y., and Purcell, E. M., *Phys. Rev.*, **94**, 630 (1954).
7. Hazlewood, C. F., Nichols, B. L., Chang, D. C., and Brown, B., *Johns Hopkins Med. J.*, **128**, 117 (1971).
8. Meiboom, S., Luz, Z., and Gill, D., *J. Chem. Phys.*, **27**, 1411 (1957).
9. Wang, J. H., *J. Chem. Phys.*, **69**, 4412 (1965).
10. Hansen, J. R., and Lawson, K. D., *Nature*, **225**, 542 (1970).
11. Woessner, D. E., *J. Phys. Chem.*, **67**, 1365 (1963).
12. Wayne, R. C., and Cotts, R. M., *Phys. Rev.*, **151**, 264 (1966).
13. Robertson, B., *Phys. Rev.*, **151**, 273 (1966).
14. Glasel, J. A., *Nature*, **227**, 704 (1970).
15. Hori, T., Teion Kagaku, Butsuri (Low Temp. Science), **15**, 34 (1956). English Translation No. 62, *US Army Snow, Ice and Permafrost Establishment*, Wilmette, Illinois.
16. Frank, H. S., and Wen, W. Y., *Discussions Faraday Soc.*, **24**, 133 (1957).
17. Horne, R. A., Day, A. F., Young, R. P., and Yu, N. T., *Electrochimica Acta*, **13**, 397 (1968).
18. Schultz, R. D., and Asunmaa, S. K., *Recent Progress in Surface Science*, **3**, 291 (Academic Press, New York, 1970).
19. Drost-Hansen, W., *Ind. Eng. Chem.*, **61**, 10 (1969).

20. Drost-Hansen, W., in *Chemistry of the Cell Interface* (edit. by Brown, H. D.) (Academic Press, New York, in the press).
21. Grant, E. H., *Ann. NY Acad. Sci.*, **125**, 418 (1965).
22. Hinke, J. A. M., *J. Gen. Physiol.*, **56**, 521 (1970).
23. Ling, G. N., and Negendank, W., *Physiol. Chem. and Phys.*, **2**, 15 (1970).

**Notes after publication:**

Although this is an old paper, it is a significant landmark in the development of the MRI (Magnetic Resonance Imaging) technology for medical imaging:

1. This is the first study of cellular water using the spin-echo NMR technique.
2. This study provided the first evidence that the nuclear magnetic relaxation times ($T_1$ & $T_2$) of cellular water protons are very different from that of bulk water protons.
3. This discovery allowed the development of the MRI technique to image biological tissues. Had the relaxation times of protons not different between cellular water and bulk water, it would be impossible to differentiate cells & tissues from extracellular water in the body by NMR. In that case, there would be no contrast in the MRI imaging.
4. In a later paper (PNAS, 1972), we also showed that the nuclear magnetic relaxation times ($T_1$ & $T_2$) of water protons in tumors and pre-tumors are significantly different from normal cells. This enabled us to use MRI for diagnosis and prognosis of cancer in a patient.